**Tunable on-chip optical traps for levitating particles based on single-layer metasurface**

*Chuang Sun, Hailong Pi, Kian Shen Kiang, Tiberius S. Georgescu, Jun-Yu Ou, Hendrik Ulbricht, and Jize Yan*

Chuang Sun, Hailong Pi, Kian Shen Kiang, and Jize Yan*
School of Electronics and Computer Science, University of Southampton, Southampton SO17 1BJ, UK
*J.Yan@soton.ac.uk

Tiberius S. Georgescu, Jun-Yu Ou*, Hendrik Ulbricht
School of Physics and Astronomy, University of Southampton, Southampton SO17 1BJ, UK
*bruce.ou@soton.ac.uk

**Optically levitated multiple nanoparticles has emerged as a platform for studying complex fundamental physics such as non-equilibrium phenomena, quantum entanglement, and light-matter interaction, which could be applied for sensing weak forces and torques with high sensitivity and accuracy. An optical trapping landscape of increased complexity is needed to engineer the interaction between levitated particles beyond the single harmonic trap. However, existing platforms based on spatial light modulators for studying interactions between levitated particles suffered from low efficiency, instability at focal points, the complexity of optical systems, and the scalability for sensing applications. Here, we experimentally demonstrated that a metasurface which forms two diffraction-limited focal points with a high numerical aperture (~0.9) and high efficiency (31%) can generate tunable optical potential wells without any intensity fluctuations. A bistable potential and double potential wells were observed in the experiment by varying the focal points' distance, and two nanoparticles were levitated in double potential wells for hours, which could be used for investigating the levitated particles' nonlinear dynamics, thermal dynamics, and optical binding. This would pave the way for scaling the number of levitated optomechanical devices or realizing paralleled levitated sensors.**

# 1. Introduction

Since optical levitation was experimentally demonstrated in the air (1971) [1] and vacuum (1976) [2], it has experienced remarkable progress over the past decade. A nanoscale or microscale particle optically levitated in vacuum is well isolated from the thermal bath and molecules collision, which makes it a promising platform to study the quantum behaviour of a macroscopic object at room temperature and the first important step has been achieved by cooling the motion of the trapped particle to its quantum ground state [3-6]. Because of the absence of dissipation by conduction and physical contact, it has been predicted that the motion of the centre of mass of an optically levitated particle could obtain a mechanical quality factor (Q factor) as high as $10^{12}$ in vacuum and Q factor of $10^{10}$ has been achieved experimentally [7] which makes levitated mechanical systems a promising candidate for ultrasensitive force, torque, and acceleration sensing [8-10], and for testing fundamental physics including, the foundations of quantum mechanics and gravity, new particles and fields beyond the standard model of quantum physics, gravitational waves, dark matter, and dark energy, but also effects of the quantum vacuum such as the Casimir force [11-14]. Recently, levitating a particle in multi-stable potentials (e.g., the w-shape double-well potential) is attracting more attention because it is an excellent platform for investigations of non-equilibrium thermodynamics, nonlinear dynamics, and macroscopic quantum tunnelling [15-18].

At the same time, the levitation of multiple particles has become technically possible for studying multi-particle entanglement, nonreciprocal dynamics, and particle-particle interactions such as optical binding [14, 19-22]. The ability to trap and manipulate simultaneously multiple particles or even an array of particles in a vacuum will be of vital importance [14, 20, 21]. In general, there are two ways of trapping more than one particle at a time. The first one way is using the standing wave generated by two counter-propagating beams [22, 23], and another one way is introducing a modulator [e.g., spatial light modulator (SLM), acoustic-optics deflector (AOD), digital micromirror device (DMD)] to the optical levitation system [24, 25].

Because a 2-level blazed grating phase profile is required to split the incident laser beam in two symmetric diffraction orders in a SLM-based optical traps system, the light utilization efficiency of the SLM would be lower 30% in the wavelength of 1550nm [29]. While the transmission of an objective lens with high numerical aperture can be corrected to be around 85% in visible spectrum, that would be only around 50% in near infrared spectrum [27]. As a result, the overall efficiency of a SLM-based optical traps would be only around 5% because there is some insertion loss arising from the other bulky optical components. In addition, limited by the modulation principle of a SLM, there would be continuous fluctuation in the intensity of the two focal points. In an AOD-based optical traps, because of the diffraction loss and the objective loss, light utilization efficiency is low as well. In addition, the optical intensity of each trap would be continuous fluctuation as well because of the instability of RF frequency and power. What's more, the optical frequency of each trap is different. Both the intensity fluctuation and optical frequency difference highly suppress the interactions between trapped particles [30].

Remarkably, it has been shown that transferring spin angular momentum (SAM) carried by a circularly polarized (CP) trapping beam [26] to a trapped particle can be used to rotate the nanoparticle at high speed. It has been further shown a high-speed rotation effectively removes particle's structural anisotropies arising from fabrication limitations and prevents motion instabilities [27]. This then requires tuning both focal points to be circularly polarized (CP) by inserting polarization modulation elements between the focal plane and the modulator. As a result, the optical levitation system becomes bulky, complicated, and difficult to align, making modulator-based systems inconvenient for practical applications [19-21] and the scaling of the number of devices [28]. The low efficiency and intensity instability are further disadvantages of using a modulator-based optical traps.

To pave the way for scaling the number of levitated optomechanical devices or realising of paralleled levitated sensors as well as improving the efficiency, it is essential to miniaturize and simplify the levitation system and one option appears to be the realisation of levitated sensors on the chip. Benefiting from the advancement of nanofabricated silicon photonic devices such as metasurface acting on lights properties in the near-field regime [31 - 40], the metasurface provides an opportunity for a new platform for mass produced optical levitation systems with high efficiency. Recently, Li's group demonstrated optical levitation utilizing a high-NA metasurface [34]. However, there is no experimental report on simultaneously levitating two particles with a metasurface-based optical levitation system.

Here, we propose a scalable single chip with tunable optical traps for realizing bistable potential well and double potential wells with high NA, high efficiency, and no intensity fluctuation at focal points, which can be extended to multiple-particle traps benefiting from the breakthrough in the limitation of polarization-multiplexing metasurface [41], shown in **Figure 1**. Importantly, the focal points are inherently circularly polarized and support trapping particles with high-speed spin modes. We experimentally obtained two near-diffraction-limited focal points at the wavelength of 1550nm with a high numerical aperture of 0.9, and high light utilization efficiency of 31%, and our experiments demonstrate that the distance between two focal points can be accurately tailored, and relative intensity between two traps can be dynamically and continuously tuned by changing the polarization state of the incident laser beam. Therefore, the particles levitated in each focal point would experience tunable optical trapping potential. While the distance between two focal points is close enough, the focal points will combine and can provide a bistable potential well. Finally, we experimentally show the metasurface's ability to construct an on-chip optical levitation system with tunable potential wells.

## 2. Metasurface Design and characterization

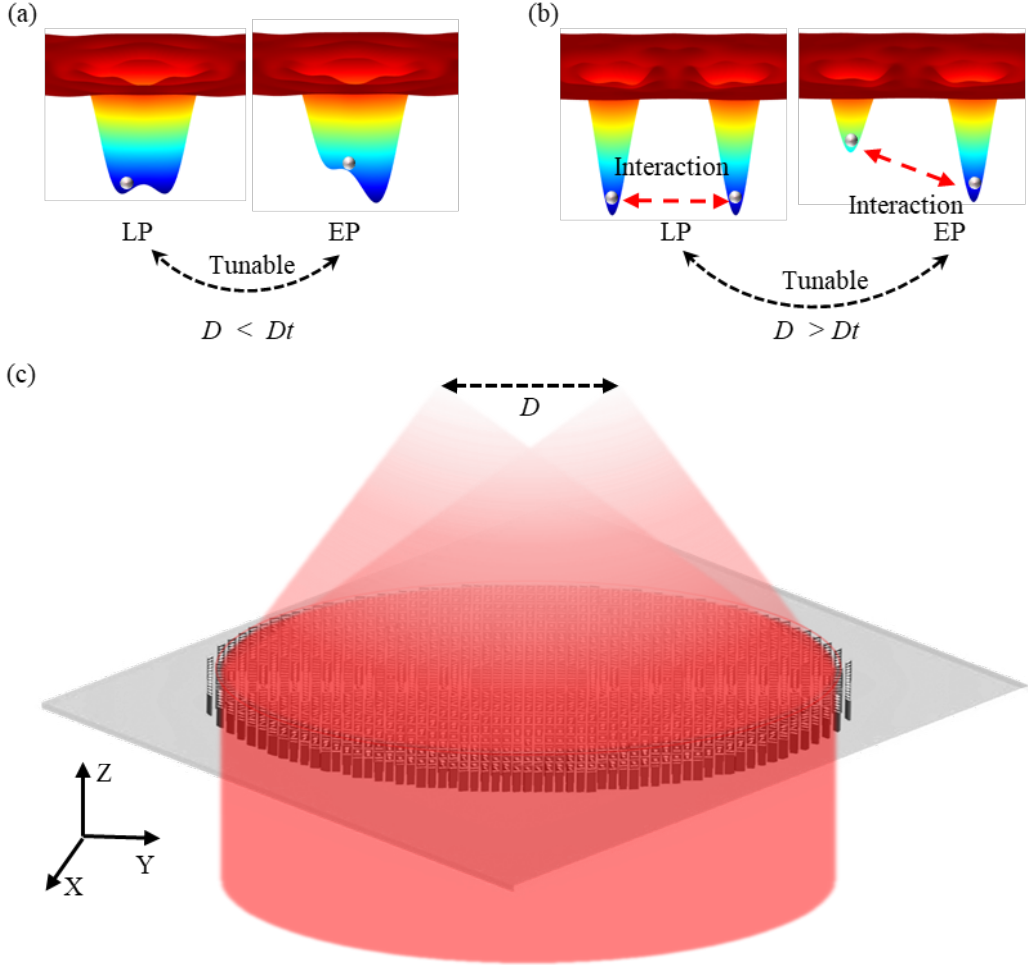

**Figure 1.** (a) A tunable bistable potential well when the distance $D$ is smaller than a threshold $D_t$ for non-equilibrium thermodynamics or nonlinear dynamics. (b) A tunable two potential wells when the distance $D$ is larger than a threshold $D_t$ for optical binding application. (c) One-chip optical levitation system based on a single-layer metasurface. [See supplementary **Figure S2d** for the determination of $D_t$]

We design the metasurface for controlled double focal points by combining the spin-multiplexing [33, 37] of the Pancharatnam-Berry-phase and an out-of-plane focusing phase profile [38]. In our metasurface, one incident laser beam could be directly focused to two or more circularly polarized focal points with high NA, as shown in **Figure 1**, which would effectively reduce the loss and instability arising from the optical components in an SLM-based optical binding system. According to the design principle and procedures described in supplement material, distance $D = 2f\tan(\alpha)$ between two focal points could be easily tailored by setting the focal length $f$ and the tilt angle $\alpha$ (**Figure S1c** in supplement material S1), which is firstly confirmed in simulation (Supplement material S2). As shown in **Figure 1**, by tailoring the distance $D$, the potential well could be shaped from the bistable potential well (**Figure 1a**) for a single particle to separately double potential wells (**Figure 1b**) for the optical binding of two particles. As the two focal points respectively correspond to the left-circularly-polarized (LCP) and right-circularly-polarized (RCP) components, the relative intensity between two focal points can be tuned by manipulating the polarization of the incident laser beam (Details see supplement material S3).

As discussed in supplement material S2 and shown in **Figure 1b**, the single layer metasurface can provide a bistable optical potential which can stably trap a single particle when the distance $D$ is smaller than $D_t$. Benefiting from the tunability of the relative depth of the double-well potential, the metasurface can work for studying the nonlinear and thermal dynamics of one levitated particle.

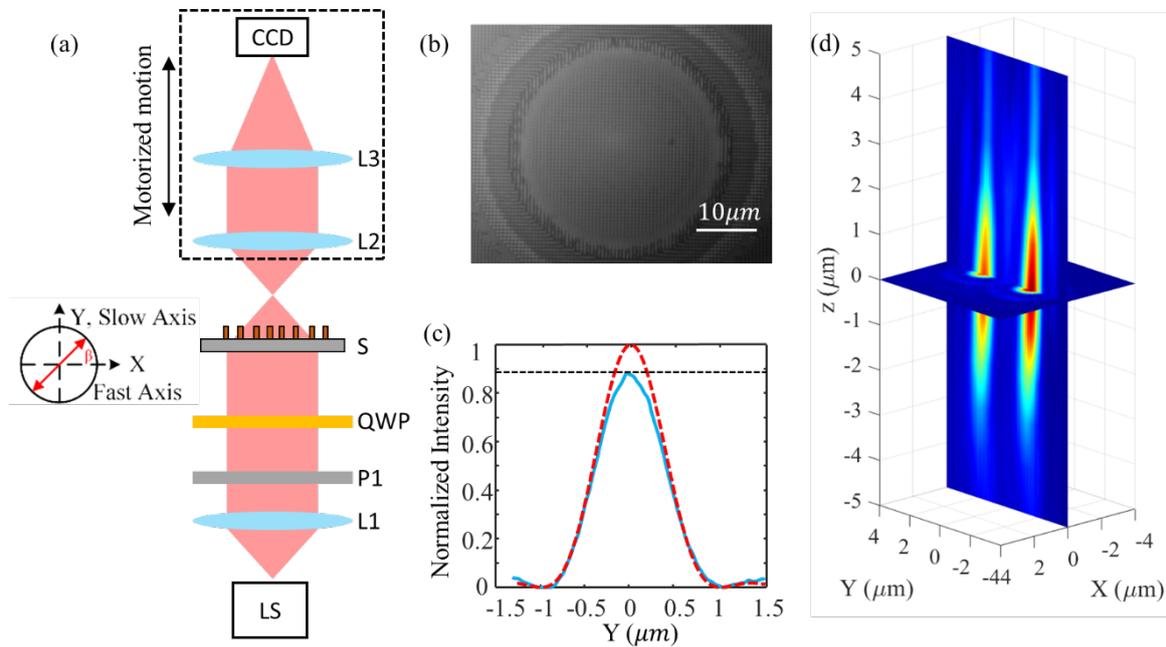

**Figure 2.** (a) Optical characterization setup configuration. LS: 1550nm fibre laser source; L1, fibre collimator; P1: polarizer; QWP: quarter waveplate; S: metalens sample; L2: NA=0.9 100X Nikon objective lens; L3: tube lens; CCD: InGaAs-based; (b) Central image of the metalens; (c) Strehl ratio of samples. Red dashed curve: diffraction limitation corresponding to 0.9; Blue line: measured intensity at focal plane; (d) focusing process near the focal plane.

In the experiment, a serial of metasurface samples with the same radius (600μm) and focal length (300μm) (i.e., NA=0.9) are fabricated and measured in our lab. To measure the optical performance of fabricated samples, an optical setup is built in our lab. As shown in **Figure 2a**, a laser beam from a fibre laser is first collimated by lens L1 to fully illuminate the sample (S). Note that a polarizer P1 and quarter waveplate (QWP) are placed between the beam expander and the sample to tune the polarization of the incident laser beam. Then, the laser beam from the sample is collected by an objective lens (L2) and imaged to an InGaAs-based CCD by a tube lens (L3). The imaging system from L2 to CCD is motorized for precisely measuring its focal length.

For measuring the focal length and focal point's intensity distribution, the objective lens L2 is firstly focused on our metalens (**Figure 2b**). Then, the imaging system is driven to move away from the metalens' surface for finding the focal plane and measuring the focal points' intensity distribution. In experiment, the distance (i.e., the focal length) from the our metalens to the focal plane is measured to be 300μm which precisely matches with our design value of 300μm. **Figure 2d** shows the focusing process from the 5μm (Z in negative value) in front of the focal plane to the 5μm (Z in positive value) behind the focal plane.

For measuring the light utlization efficiency of our metalens, the incident laser power $P_0$ is firstly measured at the rear surface of the metalens, and then the laser power $P_1$ arrived at focal plane is measured. The light utlization efficiency of $P_1/P_0$ is measured to be 31%. Note that the laser beam width is adjusted to match with the diameter (1.2mm) of our metalens, and the CCD is changed to be a power meter for measuring the powers. Notably, the overall light utilization efficiency (laser power at focal point/incident laser power) of a metasurface-based optical levitation system is the metasurface's light utilization efficiency as there are no other components in the optical path. Therefore, by using the metasurface, the overall light utilization efficiency of an optical levitation system with tunable two optical traps could be improved to 31% from 5%. In addition, the light utlization efficiency of the metasurface could be further improved by optimising the structural parameters of each cell, the materials, and the fabrication processing [42].

To furtherly evaluate the focusing performance of our metalens, the Strehl ratio ($S$) of the sample is measured. As shown in **Figure 2c**, the Strehl ratio of our sample is 0.89. According to the relationship $S = \exp[-(2\pi\sigma)^2]$ between the Strehl ration and the RMS wavefront error $\sigma$, the RMS wavefront error is calculated to be $0.054\lambda$ (wavelength $\lambda = 1550$nm), which meets the diffraction limitation criterion.

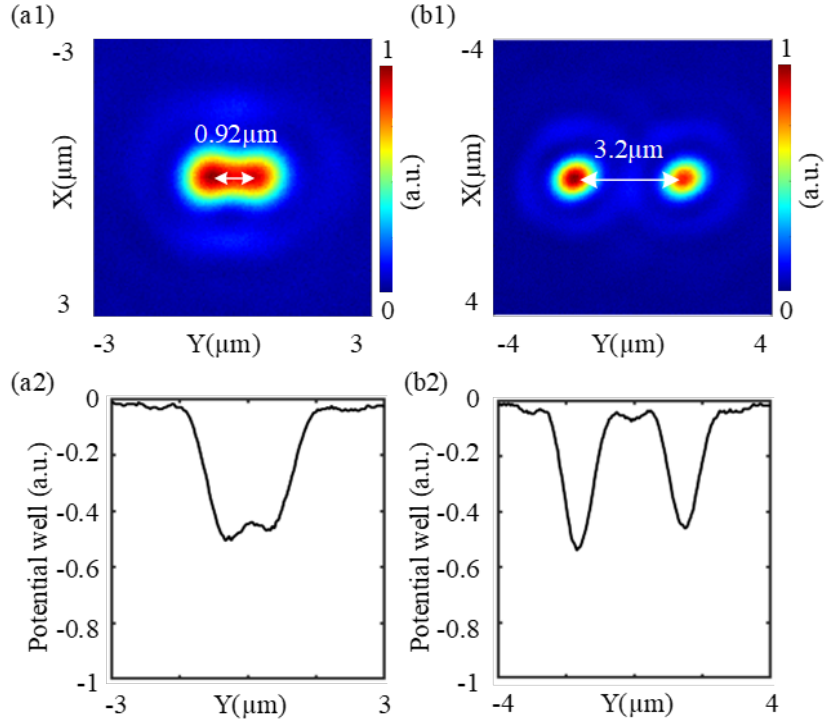

**Figure 3.** Relative optical intensity (1st row) and potential well (2nd row) in the focal plane of the metasurface with a target distance $D$ of 0.9μm (a) and 3.2μm (b). Incident laser beam is nearly LP.

F**igure 3** shows the optical intensity and relatively optical potential distributions of two samples (**Figure 3a** is corresponding to $D = 0.92$μm, and **Figure 3b** corresponding to $D = 3.2$μm) when the incident laser beam is linearly polarized. Note that as the incident laser beam is not perfectly LP, the relative intensity of two focal points is not precisely identical. As shown in **Figures 3a1** and **3b1**, the real distance of two samples is measured to be 0.92μm and 3.2μm, which are almost equal to the target values 0.9μm and 3.15μm. Comparing the **Figure 3a** and **3b**, the central depth of the 0.92μm sample is slow, while that of the 3.2μm sample is high enough for isolating the potential into two wells and trapping 2 particles for simultaneously levitating two particles at each focal point.

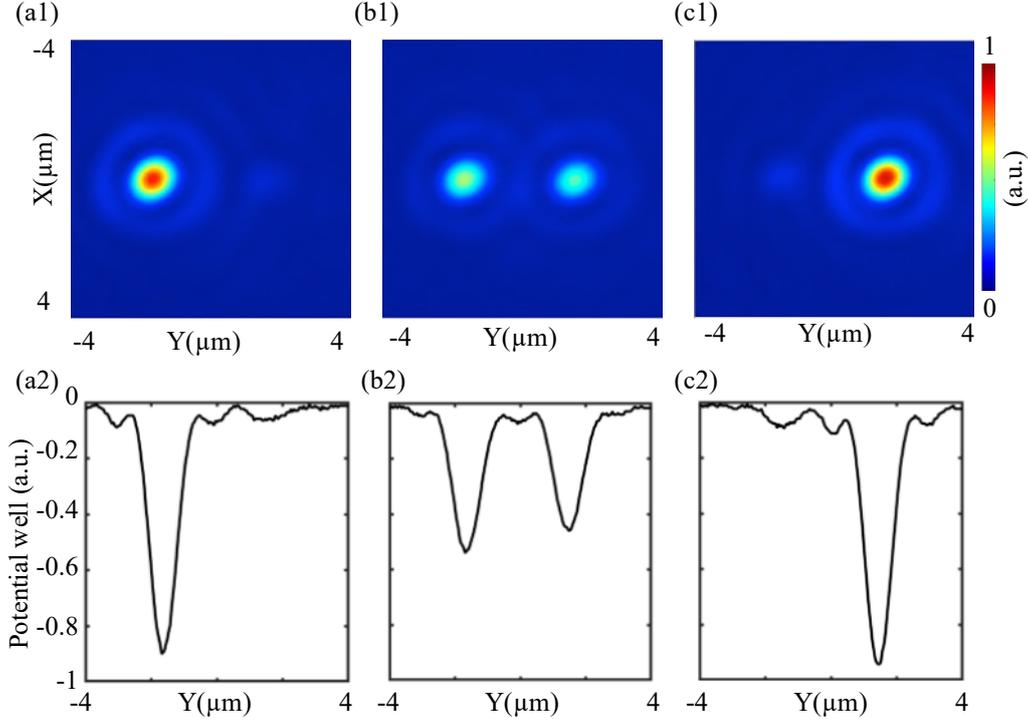

**Figure 4.** Relative optical intensity (1st row) and potential well (2nd row) in the focal plane of the metasurface with a target distance D of 3.2μm when the incident laser beam is LCP (a), LP (b), and RCP (c).

The incident laser beam is tuned to be LCP and RCP for determining the focusing performance of the metasurface and demonstrating the tunability of two focal points' relative intensity and potential wells. The polarization is manipulated by changing the angle β of a quarter-waveplate (QWP) in the experiment (Figure 2a). When β = -45°, 0°, and 45°, focal field's intensity distribution of the 3.2μm sample with is measured and shown in **Figure 4**. By changing the β from -45° to 45°, the left focal point's intensity goes down, and the right one's intensity goes up. When β = 0°, two focal points with identical intensity are obtained **Figure 4b**. It indicates that the relative intensity could be well controlled by the rotation angle β of a QWP. Supplementary movie S1 shows the evolution process in detail. As the optical trapping potential well's depth is linearly scaled with the intensity of the focal point (**Equation S3)**, two tunable trapping potential wells are obtained in the experiment as indicated by the second row in **Figure 4**. According to the intensity distribution shown in **Figure 4**, the full width of half maximum (FWHM) of the focal points are measured to be 930nm, which is corresponding to a NA of 0.9.

### 3. Optical Levitation Experiment

The optical potential wells shows that the metasurface can stably trap particles at both focal points, shown in **Figure S2**, and a setup is built to verify the optical levitation ability of the fabricated metasurface in free space. As the red dashed line has shown (**Figure 5a**), the collimated 1550nm laser beam is introduced into the rear surface of the fabricated metasurface and then is collected by an objective lens (NA=0.9) for alignment and imaging of the focal points by a 1500-1600nm NIR CCD (fluorescence based). The 1550nm laser beam is for trapping particles. In addition, a fibre-based polarization controller is introduced to manipulate the beam's polarization. For imaging the levitated particles, as the green dashed line shown (**Figure 5a**), a green laser beam (520nm) is glared injected to the focal plane of the metasurface, and the green laser would be scattered by the levitated particles, collected by the objective lens, and imaged by a visible CMOS camera. A long-pass dichroic mirror (550nm cut-on wavelength) is placed behind the objective lens for split the 1550nm trapping laser and the 520nm detection laser beams.

In experiment, the optical path of the 1550nm laser is firstly aligned by checking the focal point's images via the 1550nm imaging system. A brightest image with circular profile (**Figures 5b1** and **5c1**] can be obtained when the optical path is aligned well enough. At this time, the focal planes of metasurface and the objective lens are coincided at a same plane at 1550nm. Otherwise, the image would be blurred by the misalignment (e.g., coma and defocus aberrations). Then, the metasurface is moved forward to the objective lens by $300 \mu m$ (i.e., the focal length of the metasurface) via a precision translation stage, and the functional surface (i.e., the former surface) of metasurface is moved to the focal plane of the objective lens, which plays an important role for aligning the optical path of 520nm laser.

The 520nm laser beam is glared injected (i.e., the incident angle is almost equal to 90 degree) to the functional surface of the metasurface, and the nanostructure of the functional surface can be imaged by the Thorlabs CMOS camera because of the scattering effect. Next, the position of the metasurface is slightly translated from the 1550nm focal plane for obtaining a best image because of the chromatic aberration of the objective lens. The functional surface of the metasurface is located at the focal plane of the objective lens at 520nm. Finally, the metasurface is moved backward from the objective lens by $300 \mu m$, and the 1550nm focal plane of metasurface is moved to the 520nm focal plane of the objective lens, which is essential for imaging the levitate particles (**Figures 5b2** and **5c2**).

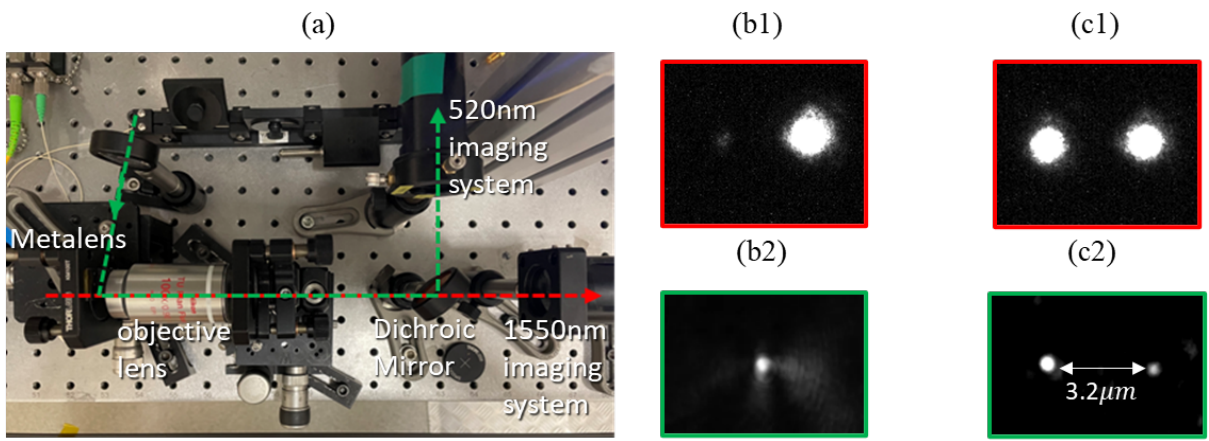

**Figure 5.** (a) Optical levitation setup. (b) and (c) are the images of the metasurface's focal points (1st row) and levitated particles (2nd row) when the incident laser beam is RCP and LP, respectively.

The 1550nm trapping laser beam (180mW) is firstly manipulated to be RCP, and a single focal point is measured by the 1550nm imaging system. Then, the solution with 100nm nanoparticles is sprayed to the region between metasurface and the objective lens via a nebulizer. Before a particle is trapped by the focal point, the image captured by the 520nm imaging system is fully dark, and then a bright spot (**Figure 5b2**) appears on the image demonstrating that a particle is trapped by the focal point. The trapping process can be more clearly seen from the supplementary movie S2, which is recorded from no particle is trapped to one particle being trapped for a long time and finally flying away. From the movie, we could clearly see the motion of the trapped particle. As this experiment is done in free space, it means that the particle is stably levitated by the focal point. In another experiment, a particle is stably levitated in air for several hours.

To verify the levitation ability of 2 particles at a same time, the 1550nm trapping laser beam (180mW) is tuned to be LP, and two focal points (**Figure 5c1**) are captured by the 1550nm imaging system correspondingly. As the particles is loading via a nebulizer, it takes some time for particles to be trapped by the focal points since they are sprayed out. As shown in supplementary movie S3, a particle is firstly trapped by the left focal point (marked by a red letter L), and another particle is then trapped by the right focal point (marked by a red letter R). **Figure 5c2** shows a snapshot of the movie S3 where two particles are trapped at a same time, which demonstrates that our metasurface-based optical levitation system could be used for simultaneously levitating two particles and studying optical binding in experiment.

We also admit that it is difficult to simultaneously trap two particles at the focal plane as shown in **Figure 5c2**. In general, only one particle can be trapped by one of the two focal points at one time as shown in the supplementary movie S4. In addition, the trapped particles in **Figure 5c2** are not identical in shape and size which makes it difficult to directly studying the dynamics of levitated particles. It is well known that these two problems are caused by the nebulizer-based loading method. In future, we'll focus on optimising the particle loading method and moving the setup into a vacuum chamber to explore the optical levitation dynamics ranging from nonlinear duffing equation to optical binding.

## 4. Conclusion

In this paper, we proposed a scalable on-chip platform for realizing tuanble optical potential wells of two trapped particles via a metasurface. Based on the metasurface, one incident laser beam can be directly focused to two diffracted-limitation focal points with high light utilization efficiency (31%) and high NA (0.9). Benefiting from the application of this metasurface, there are no other optical components in the whole system, which results that the overall light utilization efficiency of one metasurface-based optical levitation system is improved to 31%. In addition, we conceive that a higher efficiency could be obtained by optimising the structural parameters and fabrication processing. We demonstrated that the distance and relative intensity between two focal points as well as the relative potential wells could be well tailored and finely controlled in both simulation and experiment. We experimentally demonstrated that the fabricate metasurface could be used for stably levitating one or two particles for several hour in experiment, while the particle's loading method needs to be furtherly improved.

**Supporting Information**
Supporting Information is available from the Wiley Online Library or from the author.

**Acknowledgements**
PhD studentship from the Chinese Scholarship Council is acknowledged.
This work is supported by the UK's Engineering and Physical Sciences Research Council (project EP/V000624/1). Further, we would like to thank for support the UK funding agency EPSRC under grants EP/W007444/1, EP/V035975/1, and EP/X009491/1, the Leverhulme Trust (RPG-2022-57), the EU Horizon 2020 FET-Open project *TeQ* (766900), the EU Horizon Europe EIC Pathfinder project *QuCoM* under the UKRI Innovate UK compensation scheme (GA:10032223), and the QuantERA grant LEMAQUME, funded by the QuantERA II ERA-NET Cofund in Quantum Technologies implemented within the EU Horizon 2020 Programme

**Data Availability Statement**
The data from this paper can be obtained from the University of Southampton ePrints research repository: https://doi.org/10.5258/SOTON/D2470. All data needed to evaluate the conclusions in the paper are present in the paper and/or the Supplementary Materials.

# Supplemental material: Tunable on-chip optical traps for levitating particles based on single-layer metasurface

*Chuang Sun, Hailong Pi, Kian Shen Kiang, Tiberius S. Georgescu, Jun-Yu Ou, Hendrik Ulbricht, and Jize Yan*

Chuang Sun, Hailong Pi, Kian Shen Kiang, and Jize Yan*
School of Electronics and Computer Science, University of Southampton, Southampton SO17 1BJ, UK
*J.Yan@soton.ac.uk

Tiberius S. Georgescu, Jun-Yu Ou*, Hendrik Ulbricht
Department of Physics and Astronomy, University of Southampton, Southampton SO17 1BJ, UK
*bruce.ou@soton.ac.uk

**S1. Principle and procedure of designing metasurface**

In general, a metasurface is consisting of an array of nanopillars or nanofins with fixed height and period [H and P in Fig. 1(b)] and imparts the target phase profile onto the incident laser beam. It has been proved that a metasurface can simultaneously impart independent phase profiles to the left-hand circularly polarized [LCP, i.e., $= \begin{bmatrix} 1 \\ i \end{bmatrix}$] and right-hand circularly polarized [RCP, i.e., $|R\rangle = \begin{bmatrix} 1 \\ -i \end{bmatrix}$] laser beams [s1], when the nanofins [Fig. S1(a)] act as a halfwave plate.

In addition, a linearly polarized (LP) laser beam $\begin{bmatrix} 1 \\ 0 \end{bmatrix}$ is inherently regarded as the sum of two orthogonal components LCP $\begin{bmatrix} 1 \\ i \end{bmatrix}$ and RCP $\begin{bmatrix} 1 \\ -i \end{bmatrix}$ [s1]. Therefore, a metasurface based on nanofins will focus the LCP and RCP components in one single linearly polarized beam into two focal points by imparting two different focusing phase profiles onto each component.

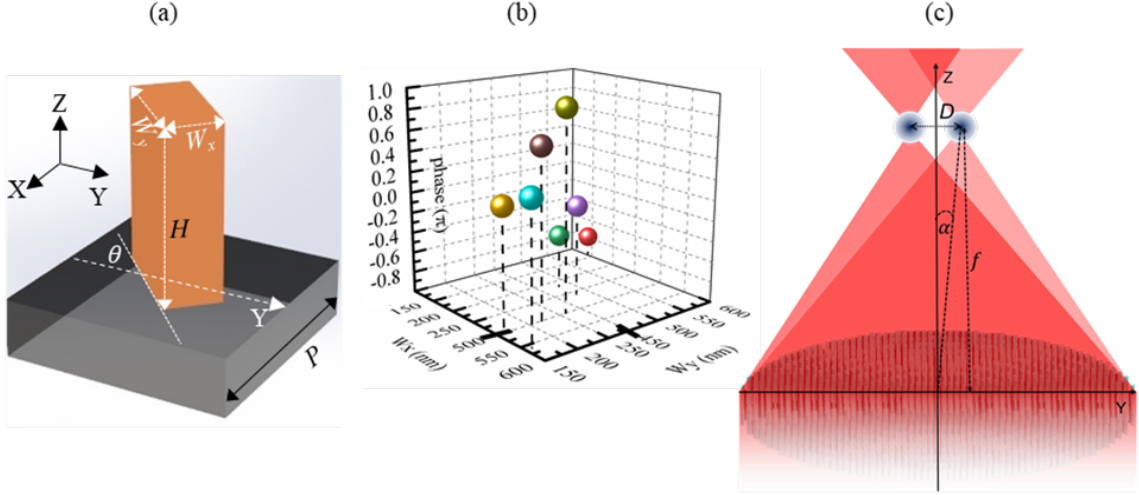

**Figure S1.** (a) Nanofin and structural parameter; (b) Selected 8 nanofins size; (c) Geometrical parameters of two focal points.

As the nanofins [Fig. S1(a)] are anisotropic in shape, the imparted phase along the X-direction ($\varphi_x$) and Y-direction ($\varphi_x$) will be different and can be modulated by varying the dimension ($W_x$ and $W_y$). When $|\varphi_x - \varphi_y| = \pi$, the nanofins can be regarded as a halfwave plate and two different phases $\varphi_L$ and $\varphi_R$ [Eq. (S1)] will be separately encoded onto the incident LCP and RCP components,

$$\begin{cases} \varphi_L = \varphi_x + 2\theta, \\ \varphi_R = \varphi_x - 2\theta; \end{cases} \tag{S1}$$

where $\theta$ is the rotation angle of the nanofins, as shown in Fig. S1(a). Therefore, $\varphi_x$ should be able to vary from $-\pi$ to $\pi$ to realize a full $2\pi$ phase coverage for both $\varphi_L$ and $\varphi_R$. As shown in Fig. S1(b), for target wavelength $\lambda = 1550$ nm, by fixing H and P as 800 nm and 650 nm, a group of nanofins with different $W_x$ and $W_y$ are selected to achieve $\varphi_x$ variation from $-\pi$ to $\pi$ and keep $|\varphi_x - \varphi_y| = \pi$.

For focusing the LCP and RCP components into two focal points at a close distance in a same focal plane, the phase profiles $\varphi_L$ and $\varphi_R$ for LCP and RCP can be expressed as Eq. (S2) [s2], where ($r_p, \theta_p$) is the polar coordinate of the nanofin located at (x,y) on the metasurface, $f$ and $\alpha$ are the target focal length and tilt angle, as shown in Fig. S1(c). As a result, the distance D between two focal points could be expressed as $D = 2f\tan(\alpha)$ and be tailored by tuning the target $\alpha$ for a given $f$ or NA, where $NA = R/\sqrt{R^2 + f^2}$ and R is the radius of the whole metasurface.

$$\begin{cases} \varphi_L(r_p, \theta_p) = \frac{2\pi}{\lambda}[f - \sqrt{f^2 + r_p^2 - 2r_p f \sin(\alpha)\cos(\theta_p)}] \\ \varphi_R(r_p, \theta_p) = \frac{2\pi}{\lambda}[f - \sqrt{f^2 + r_p^2 + 2r_p f \sin(\alpha)\cos(\theta_p)}] \end{cases} \quad (S2)$$

In order to build a specific metasurface, we follow the general design procedure [Fig. S2] with the following steps:

(1) confirming the target radius R, focal length f, and incline angle $\alpha$;

(2) calculating the target phase profile $\varphi_L(r_p, \theta_p)$ and $\varphi_R(r_p, \theta_p)$ based on Eq. (S2);

(3) calculating the target $\varphi_x$ and rotation angle $\theta$ for each nanofin located at $(r_p, \theta_p)$;

(4) finding a suitable dimension $W_x$ and $W_y$ for each nanofin from the nanofins' library shown in Fig. S1(b).

(5) The whole metasurface could be constructed.

## S2. Tailoring distance D and potential wells

As shown in Figs. S2(a1) – S2(c1), for a given $f = 5$ μm and NA=0.9, the distance D along Y direction varies from 0.7μm to 3.15μm. Note that, as shown in Fig. S2(a), the two focal points emerge into one when $\alpha = 2°$ and $D = 0.7$ μm, because the distance D of 0.7μm is smaller than the diffraction limitation (1.18μm) of each focal point. The diameter and focal length f used in simulation are 20μm and 5μm, respectively, which are much smaller than those of fabricated samples in main text. It is limited by our computer's memory.

For each distance D, the potential wells $U(y)$ experienced by a trapped SiO2 spherical particle are calculated as well to show that the potential wells $U(y)$ can be tailored for studying optical binding using Eqs. (S3) and (S4).

$$U(y) = -\frac{2\pi n_s r^3}{c}\left(\frac{m^2-1}{m^2+2}\right)I(y) \qquad (S3)$$

$$I(y) = \frac{cn_s\varepsilon_0}{2}|E(y)|^2 \qquad (S4)$$

Where $n_s = 1$ is the refraction index of surrounding environment, $r$ is the radius of the trapped particle and assumed to be 100nm, $m = 1.48$ is the relative refraction index between the SiO2 particle and the $n_s$, $c$ is the speed of light in vacuum, and the $\varepsilon_0$ is the dielectric constant in vacuum. $E$ is the electrical field amplitude of the focal plane which is simulated via Lumerical FDTD as shown in Figs. S2(a1) – S2(c1).

As shown in Figs. S2(a2) -S2(c2), the central depth $U_m$ of the potential wells goes up with the distance D increasing. According to Ashkin's criterion, the depth of the potential well should be higher than $10k_BT$ for stably trapping a particle in the focal point. $k_B$ is the Boltzmann constant, and T is the environment temperature and assumed to be 300K in our calculation. Therefore, the distance D should be large enough for obtaining two coupled optical traps. Figure S2(d) shows the relationship between the central depth $U_m$ and the distance D, which demonstrates that a $U_m > 10k_BT$ and two optical traps can be achieved when the distance D > 1.14μm ($D_t$).

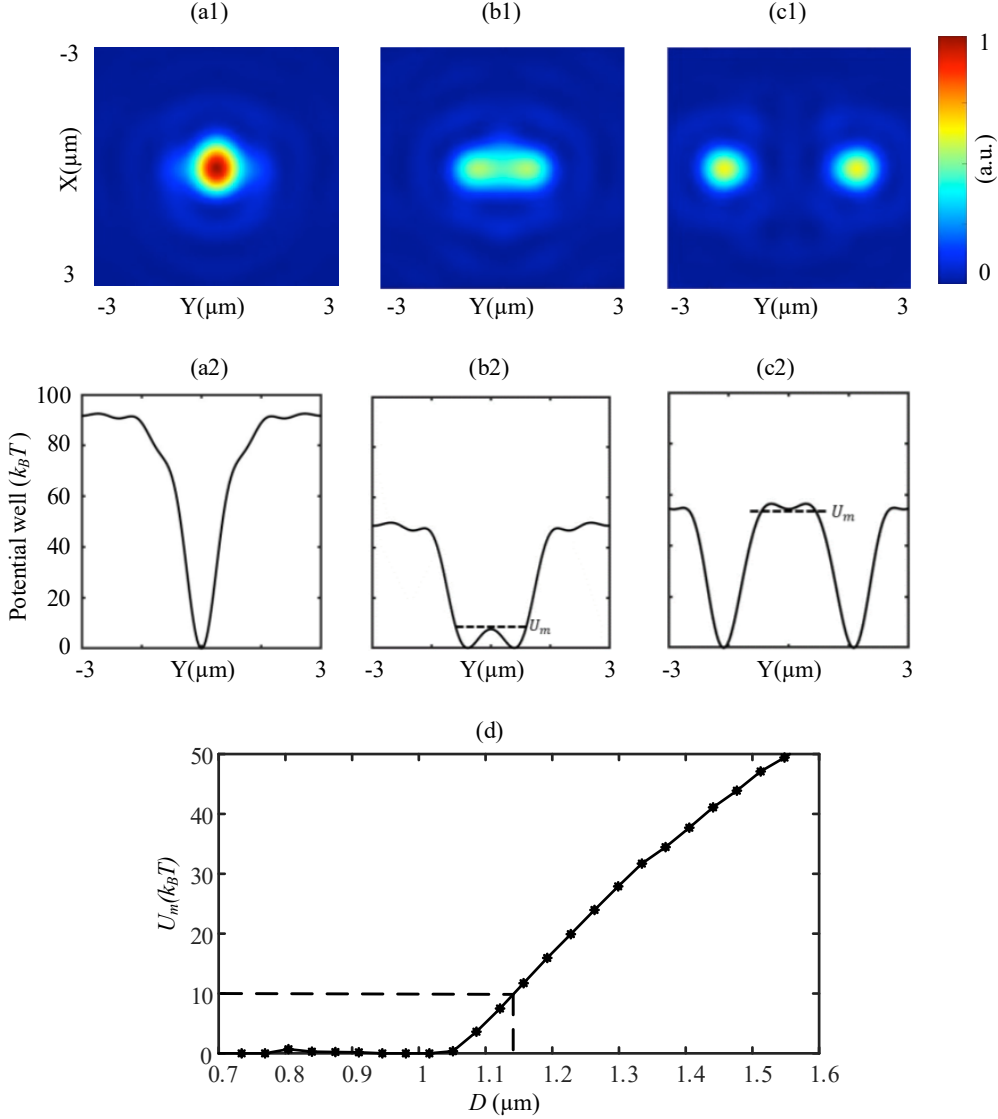

**Figure S2.** (a) –(c) are the simulated optical intensity (1st row) and potential well (2nd row) in the focal plane of the metasurface with the distance D changing from 0.7μm to 3.15μm when the incident laser beam is LP. (d) central depth of potential well vs the distance D

As shown in Figs. 1(b) and S2(b), the potential well generated by the focal points with a distance $D$ of 1.12μm (smaller than $D_t$) has a central depth $U_m$ of $7k_BT$ which cannot stably trap two particles at each focal point at a same time. As a result, the potential well shown in Fig. S2(b2) can be seen as a double-well potential or bistable optical potential well which could stable trap one single particle. Therefore, it can be used for studying nonlinear dynamics of one levitated particle.

### S3. Principle of tuning the relative intensity and potential well

Based on the coordinate system shown in Fig. S5, the Jones matrix of the incident laser beam can be expressed as $\begin{bmatrix} \cos\beta \\ -i\sin\beta \end{bmatrix}$. As a result, the amplitudes of LCP and RCP components are $[(\cos\beta - \sin\beta)/\sqrt{2}]$ and $[(\cos\beta + \sin\beta)/\sqrt{2}]$. Correspondingly, the intensities of LCP and RCP components can be expressed as $[1 - \sin(2\beta)]I_0$ and $[1 + \sin(2\beta)]I_0$, where $I_0$ means the total intensity of the incident laser beam. Therefore, the intensity of the LCP component (i.e., the left focal point) would go down from $I_0$ to zero with the angle $\beta$ changing from -45° to 45°, while the intensity of the RCP component ((i.e., the right focal point)) goes up to a maximum value $I_0$ from zero. According to the Eq. (S3), the potential well linearly scales with the intensity distribution. Therefore, the potential wells generated by the metasurface can be tuned correspondingly.

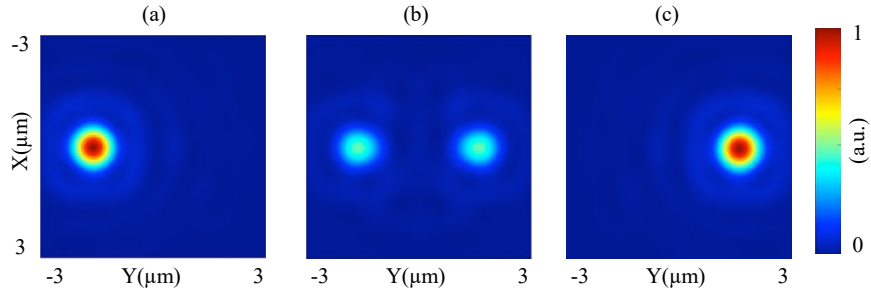

**Figure S3**. (a) –(c) are the simulated optical intensity in the focal plane of the metasurface with the distance *D* of 3.15μm when the incident laser beam is LCP (a), LP (b), and RCP (c).

As shown in Figs. 3(d), when $\beta$ = -45° (i.e., the incident laser beam is LCP), there is only the left focal point corresponding to the LCP component in the focal plane, while the RCP-related right focal point is vanished. In contrast, as shown in Fig. 3(c), when $\beta$ = 45° (i.e., the incident laser beam is RCP), there is only the right focal point corresponding to the RCP component in the focal plane, while the LCP-related left focal point is vanished. When, $\beta$ = 0° (i.e., the incident laser beam is LP), two focal points with equivalent intensity could be seen in the focal plane, which means that the relative intensity between two focal points could be tuned by rotating a QWP.

## S4. Fabrication Process

To verify the designed metalens's performance in the experiment, some samples with a diameter of 1.2mm and a target focal length of 300 $\mu m$ are fabricated in the cleanroom following the steps illustrated in Fig. S4. The metasurfaces are fabricated on a $SiO_2$ wafer. The wafer is firstly cleaned in acetone for 10mins and IPA for 10min. Then, a layer of amorph silicon (800nm in thickness) is deposited via PECVD at 300°C. ZEP520A resist is spin coated on the wafer and baked on a hot plate for 3mins at 180°C. Subsequently, the metasurface's pattern is defined by the e-beam lithography (EBL) and development process in NMP [Fig. S4 (c)]. After that, a 50nm-thick Cr layer as a hard mask is coated on the substrate by e-beam evaporation and followed by a lift-off. The patterns are transferred to the Cr layer as shown in Fig. S4 (f). Then, the wafer with the patterned Cr layer is etched by reactive ion etching (RIE) [Fig. S4 (g)]. Finally, samples are obtained after removing Cr layer via wet etch [Fig. S4 (i)].

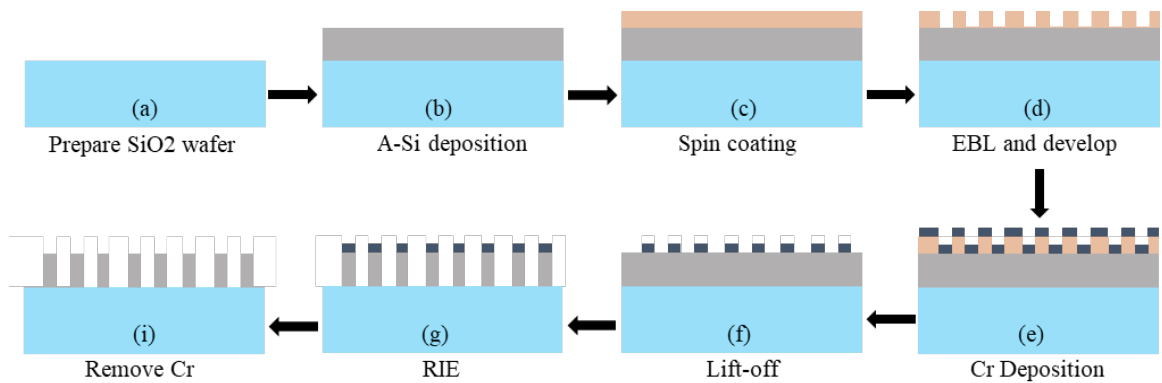

**Figure S4.** Fabrication process of metasurface.

## S5. SEM Image of fabricated samples.

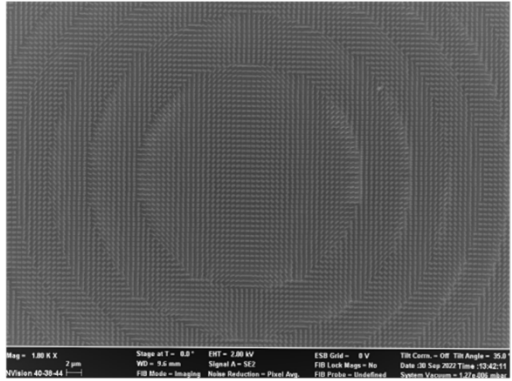

**Figure S5.** SEM image of fabricated metasurface device.

The general profile of a fabricated sample is measured by SEM, and Fig. S6 shows the central area of one sample.

Reference

[S1] T. Li, X. Xu, B. Fu, S. Wang, B. Li, Z. Wang, S. Zhu, Photonics Res. 2021, 9, 1062.

[S2] M. Khorasaninejad, W. T. Chen, A. Y. Zhu, J. Oh, R.C. Devlin, D. Rousso, F. Capasso, Nano letters 2016, 16, 4595.